\documentstyle[12pt]{article}
\begin{document}
\begin{center}
{\Large \bf 
	On possible 
	lower bounds for the direct detection rate of SUSY Dark Matter
\par }	\bigskip

{\large V.A.~Bednyakov }	\smallskip

	{\em Laboratory of Nuclear Problems,
	Joint Institute for Nuclear Research,
        Moscow region, 141980 Dubna, Russia. 
	E-mail: bedny@nusun.jinr.ru} 
\end{center}	

\begin{abstract}
	One can expect accessible lower bounds 
	for dark matter detection rate 
	due to restrictions on masses of the SUSY-partners. 
	To explore this correlation one needs 
	a new-generation large-mass detector.
 	The absolute lower bound for detection rate 
	can naturally be due to spin-dependent interaction.
	Aimed at {\em detecting}\ dark matter
	with sensitivity higher than $10^{-5}\,$event$/$day$/$kg
	an experiment should have a non-zero-spin target. 
	Perhaps, the best is to create a GENIUS-like detector 
	with both $^{73}$Ge (high spin) and $^{76}$Ge nuclei. 
\end{abstract}

	A new generation of high-sensitivity dark matter detectors,
	in particular aimed at searching for neutralinos, 
	lightest SUSY particles (LSP), 
	has been proposed (see for example, proposals GENIUS
\cite{GENIUS}, GENIUS-TF   
\cite{GENIUS-TF}, CRESST
\cite{Altmann:2001ax} and CDMS 
\cite{CDMS}).
	The question naturally arises  
	of how small the expected event rate, $R$, of the 
	LSP direct detection can be, provided the LSP is a cold dark
	matter particle (or so-called WIMP).
	The upper and lower bounds for the neutralino-nucleon cross section  
	were considered in various SUSY models 
\cite{h9701301}--\cite{Bednyakov:2001he}.
	The main goal of this paper is to 
	attract extra attention to possible 
	lower bounds for $R$, which is to be measured directly. 

	To this end the exploration of the MSSM parameter space is performed 
	at the weak scale (without any unification assumptions).
	The MSSM parameter space is determined by entries of the mass 
	matrices of neutralinos, charginos, Higgs bosons, 
	sleptons and squarks. 
	Available restrictions from cosmology
	($0.1 < \Omega_\chi h^2_0< 0.3$), rare FCNC $b\to s\gamma$ decay
	($1.0 \times 10^{-4} < {\rm B}(b\to s \gamma) < 4.2 \times 10^{-4}$), 
	accelerator SUSY searches, etc were taken into account
\cite{Bednyakov:2000vh,Bednyakov:2001he}. 

\smallskip
\begin{figure}[h] 
\begin{picture}(100,140)
\put(-5,-44){\includegraphics{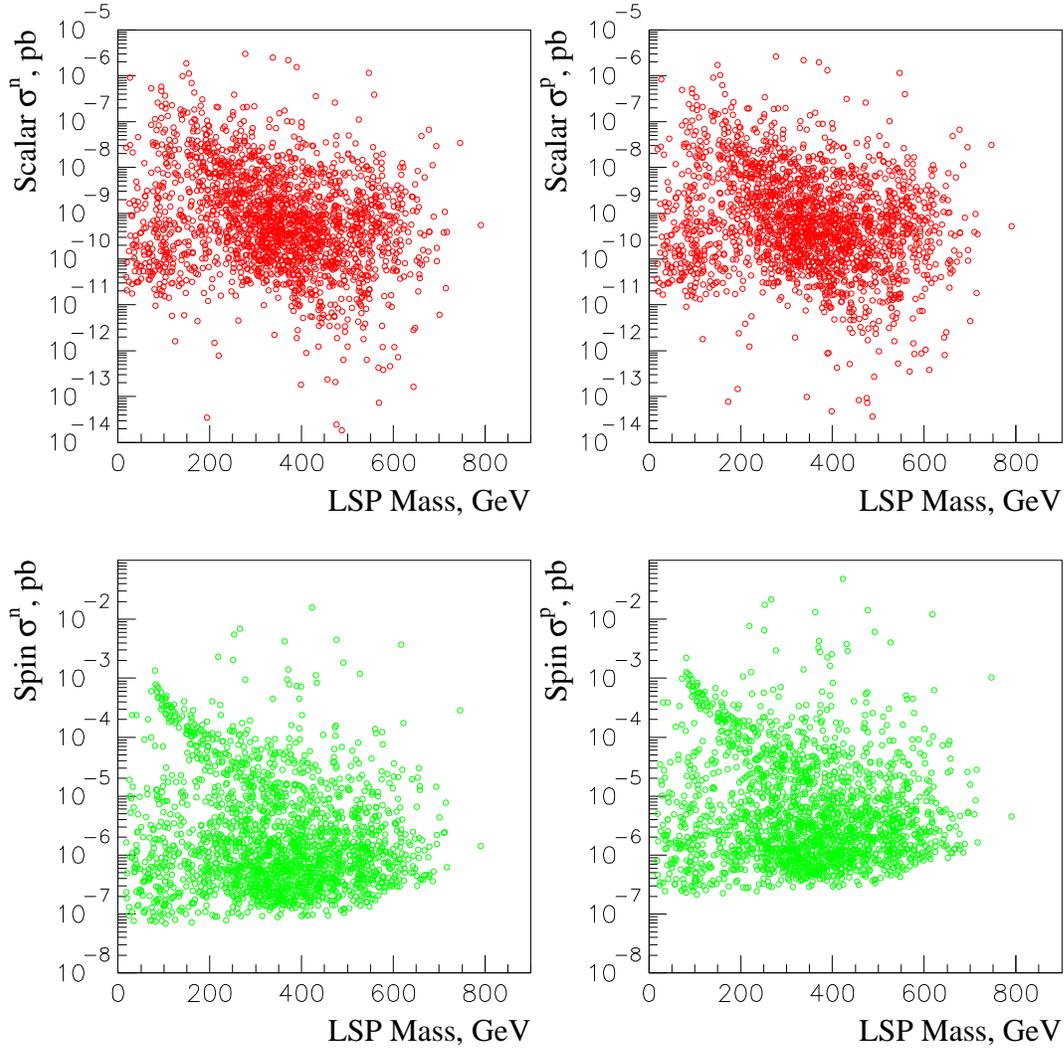}}
\end{picture}
\caption{\small 
	Cross sections of spin-dependent and spin-independent
	interactions of WIMPs with the proton and the neutron.
\label{CrossSections}}
\end{figure} 
	Scatter plots with individual cross sections of spin-dependent 
	and spin-independent (scalar) 
	interactions of LSP with the proton and the neutron
	are given in 
Fig.~\ref{CrossSections} as functions of the LSP mass.
	The different behavior of these cross sections with 
	the mass of the LSP can be seen from the plots.
	There is a 
	lower bound for the spin-dependent cross section.
	Due to the absence of a clear lower bound for the scalar cross 
	section of the WIMP-neutron interaction, 
	one can expect that the lower bound for the 
	{\em rate}\ can be  
	established by the spin-dependent interaction, 
	which in contrast to the scalar interaction is associated 
	with an about 3-order-of-magnitude larger
	WIMP-nucleon cross sections.

\begin{figure}[ht] 
\begin{picture}(100,85)
\put(20,-5){\includegraphics{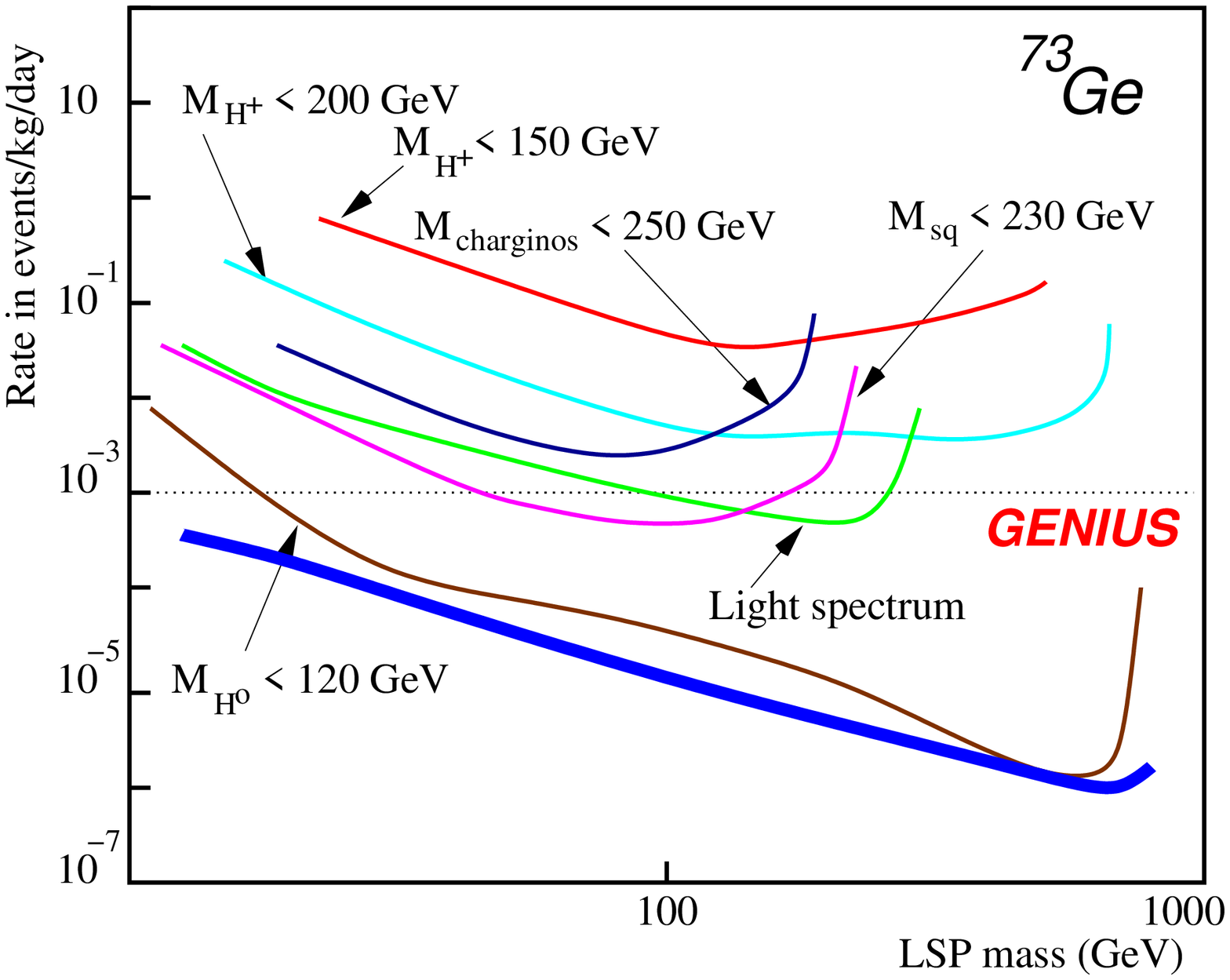}}
\end{picture}
\caption{\small
	Different lower bounds for the total  event rate in 
	$^{73}$Ge. 
	Here M$^{}_{\rm sq,\ H^+,\ H^0}$ denote masses of the squark,
	the charged Higgs boson and the light neutral CP-even Higgs boson 
	respectively.
	Heavy chargino mass is denoted as M$^{}_{\rm charginos}$.
	The thick curve corresponds to the absolute lower bound. 
	"Light spectrum" denotes the lower bound for the rate,
	obtained when all sfermion masses are lighter than 
	300--400 GeV.
	The dotted line represents the expected sensitivity
	for the direct dark matter detection with GENIUS
\protect\cite{GENIUS}.}
\label{LowBounds}
\end{figure}

	The existence of the absolute lower bound for the event rate 
(thick curve in Fig.~\ref{LowBounds})
	and the variation of the bound with the MSSM 
	parameters and masses of the SUSY particles 
\cite{Bednyakov:2000vh}
	allow one to consider prospects to searching for
	dark matter under special assumptions about 
	restricted values for the MSSM parameters and masses.

	Figure
\ref{LowBounds} gives different lower bounds for $R$ 
	obtained with extra limitations on SUSY-particle masses
\cite{Bednyakov:2000vh}.
	A restriction for the single (light) squark mass to be quite small
	(M$^{}_{\rm sq}< 230$~GeV) as well as another
	assumption that all sfermion masses do not exceed 300--400 GeV,
	put upper limits on the mass of the LSP
	and therefore do not permit $R$ to drop very deeply 
	with increasing LSP mass.
	Furthermore,  in both cases the lower bound for the
	rate is established for all allowed masses of the LSP
	at a level of $10^{-3}$ events/kg/day.     
	This value is considered as an optimistic 
	sensitivity expectation for future high-accuracy 
	detectors of dark matter, such as GENIUS 
\cite{GENIUS}.
	One can see that the mass of the light neutral CP-even 
	Higgs boson M$^{}_{\rm H^0}$ 
	has unfortunately a very poor restrictive potential. 
	The situation looks most promising when one limits the
	mass of the charged Higgs boson.
	If it happened, for instance,  
	that either the SUSY spectrum is indeed light or 
	M$^{}_{\rm H^+}<200$~GeV, in both cases at least 
	the GENIUS experiment should detect a dark matter signal.
\enlargethispage{0.7\baselineskip}

	Therefore the prospects could be very promising
	if from collider searches one would restrict M$^{}_{\rm H^+}$
	at a level of about 200~GeV. 
	The observation, due to its importance for 
	dark matter detection, could serve as a source for extra
	efforts in searching for the charged Higgs boson with colliders.
	Otherwise, non-observation of any dark matter signal
	with very sensitive dark matter detectors 
	would exclude, for example, 
	a SUSY spectrum with masses lighter than 300--400 GeV,
	charginos with masses smaller than 250 GeV
(Fig.~\ref{LowBounds}), the
	charged Higgs boson with M$^{}_{\rm H^+}<200$~GeV, etc.

	It was claimed that nuclear spin is not important 
	for detection of dark matter particles, 
	provided the detection sensitivity 
	does not exceed 0.01 event$/$day$/$kg, 
	which was considered unattainable in 1994 
\cite{h9401262}.
	Now, with new-generation detectors, the situation changes and 
	targets with spin-non-zero nuclei should again be taken into account.

\begin{figure}[h] 
\begin{picture}(100,75)
\put(-2,-78){\includegraphics{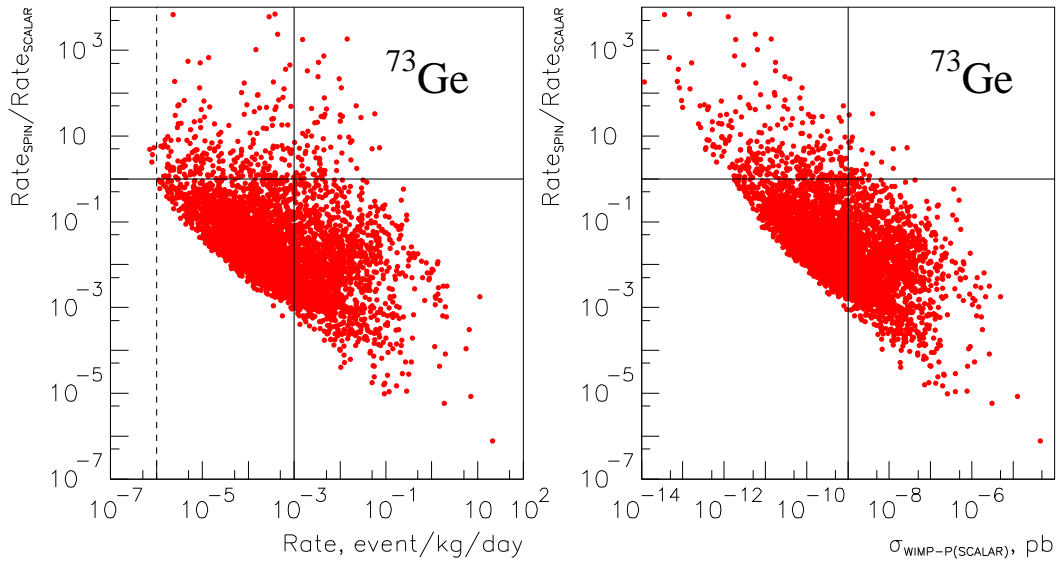}}
\end{picture}
\caption{\small 
	Ratio of the spin-dependent event rate to the   	
	spin-independent event rate in the $^{73}$Ge isotope
	(spin$\,=9/2$) as a function of the total (spin-dependent 
	plus spin-independent) event rate (left) and the scalar 
	cross section of the neutralino-proton interaction (right). 
	The solid vertical lines give the expected sensitivity of GENIUS
\protect\cite{GENIUS}. 
	In the region above the horizontal line the spin contribution 
	dominates.
\label{SpinandScalar}}	
\end{figure} 

	For any mass of the LSP one can find very large and 
	very small values for the spin-dependent to spin-independent 
	cross section (or rate) ratio.
	The spin-independent (scalar) contribution
	obviously dominates in the domain of large expected rates 
(Fig.~\ref{SpinandScalar}) 
	in the spin-non-zero germanium detector
	($R>0.1\,$event$/$day$/$kg). 
	But as soon as the total rate drops down to
	$R<0.01\,$event$/$day$/$kg or, equivalently, the
	scalar neutralino-proton cross section becomes
	smaller than $10^{-9}$--$10^{-10}\,$pb,
	the spin-dependent interaction may produce
	a rather non-negligible contribution to the total event rate.
	Moreover, if the scalar cross section further decreases
	($\sigma < 10^{-12}\,$pb), it becomes obvious that 
	the spin contribution alone saturates 
	the total rate and protects it from decreasing below
	$R\approx 10^{-6}$--$10^{-7}\,$event$/$day$/$kg
\cite{Bednyakov:2000vh}.
	With only a spinless detector  
	one can miss a signal caused by spin-dependent interaction.  
	Aimed at {detecting} dark matter
	with sensitivity higher than $10^{-5}\,$event$/$day$/$kg
	an experiment should have a spin-non-zero target. 
	Indeed, while the scalar cross sections governed 
	mostly by Higgs exchange can be rather small, 
	the spin cross section cannot be arbitrary small, 
	because the mass of the $Z$ boson, which gives the dominant 
	contribution, is well defined, provided one ignores any 
	possible fine-tuning cancellations 
\cite{Ellis:2001jd}.
\enlargethispage{0.7\baselineskip}

	Therefore, if an experiment with sensitivity 
	$10^{-5}$--$10^{-6}\,$event$/$day$/$kg
	fails to detect a dark matter signal,
	an experiment with higher sensitivity 
	should have a spin-non-zero target and  
	will be able to detect dark matter particles 
	only due to the spin neutralino-quark interaction.
	In this situation, it seems 
	the best to create a huge GENIUS-like detector 
	with both $^{73}$Ge (high spin) and $^{76}$Ge (spinless) isotopes.

\smallskip
	The author thanks Prof. H.V. Klapdor-Kleingrothaus and 
	RFBR (Grant 00-02--17587) for support. 
 

\end{document}